\documentclass[conference]{IEEEtran}

\usepackage{cite}
\usepackage{amsmath,amssymb,amsfonts}
\usepackage{algorithmic}
\usepackage{graphicx}
\usepackage{textcomp}
\usepackage{xcolor}
\def\BibTeX{{\rm B\kern-.05em{\sc i\kern-.025em b}\kern-.08em
    T\kern-.1667em\lower.7ex\hbox{E}\kern-.125emX}}

\usepackage[T1]{fontenc}

\usepackage{scalerel}

\usepackage{siunitx}
\usepackage{tikz}
\usepackage{pgfplots}
\pgfplotsset{compat=newest}
\usetikzlibrary{calc,intersections,svg.path,pgfplots.external,arrows.meta,backgrounds,fit}
\usepgfplotslibrary{statistics}

\usepackage{array}	
\usepackage{booktabs}	
\usepackage{dcolumn}	
\usepackage{multirow}
\usepackage{multicol}

\usepackage{marginnote}
\newcommand{\TODO}[1]{}

\newcolumntype{H}{>{\scriptsize}c}

\usepackage[labelformat=simple,labelfont=bf,font=footnotesize,singlelinecheck=true]{subcaption}	

\DeclareCaptionLabelSeparator{periodspace}{.\quad}
\definecolor{orcidlogocol}{HTML}{A6CE39}
\tikzset{
  orcidlogo/.pic={
    \fill[orcidlogocol] svg{M256,128c0,70.7-57.3,128-128,128C57.3,256,0,198.7,0,128C0,57.3,57.3,0,128,0C198.7,0,256,57.3,256,128z};
    \fill[white] svg{M86.3,186.2H70.9V79.1h15.4v48.4V186.2z}
                 svg{M108.9,79.1h41.6c39.6,0,57,28.3,57,53.6c0,27.5-21.5,53.6-56.8,53.6h-41.8V79.1z M124.3,172.4h24.5c34.9,0,42.9-26.5,42.9-39.7c0-21.5-13.7-39.7-43.7-39.7h-23.7V172.4z}
                 svg{M88.7,56.8c0,5.5-4.5,10.1-10.1,10.1c-5.6,0-10.1-4.6-10.1-10.1c0-5.6,4.5-10.1,10.1-10.1C84.2,46.7,88.7,51.3,88.7,56.8z};
  }
}

\newcommand\orcidlink[1]{\href{https://orcid.org/#1}{\mbox{\scalerel*{
\begin{tikzpicture}[yscale=-1,transform shape]
\pic{orcidlogo};
\end{tikzpicture}
}{|}}}}

\usepackage{hyperref}

\hypersetup{%
  pdfauthor={M. F. Schiffner},%
  pdftitle={Frequency-Dependent F-Numbers Suppress Grating Lobes and Improve the Lateral Resolution in Line-by-Line Scanning},%
  pdfsubject={Lehrstuhl für Medizintechnik, Ruhr-Universität Bochum},%
  pdfkeywords={%
    beamforming, %
    dynamic aperture, %
    F-number, %
    frequency dependence, %
    line-by-line scanning, %
    progressive scanning%
  },%
  colorlinks=false,%
  hidelinks,
  linktocpage=true,%
  bookmarksopenlevel=1%
}
\usepackage{breakurl}

\usepackage{acronym} 

\usepackage{cleveref}	  			

\crefname{chapter}{Chapter}{Chapters}		
\Crefname{chapter}{Chapter}{Chapters}		
\crefname{appendix}{Appendix}{Appendices}
\Crefname{appendix}{Appendix}{Appendices}
\crefname{section}{Sect.}{Sections}
\Crefname{section}{Section}{Sections}
\crefname{subsection}{Sect.}{Sections}
\Crefname{subsection}{Section}{Sections}
\crefname{subsubsection}{Sect.}{Sections}
\Crefname{subsubsection}{Section}{Sections}

\crefname{figure}{Fig.}{Figs.}
\Crefname{figure}{Figure}{Figures}

\crefname{table}{Table}{Tables}
\Crefname{table}{Table}{Tables}

\newcommand{\name}[1]{#1}

\definecolor{RUBBLUE}{cmyk}{1.000,0.500,0.000,0.600}	
\definecolor{RUBGREEN}{cmyk}{0.500,0.000,1.000,0.000}	
\definecolor{RUBGRAY}{cmyk}{0.030,0.030,0.030,0.100}	

\definecolor{RUBBLUE_RGB}{rgb}{0.000,0.208,0.377}
\definecolor{RUBGREEN_RGB}{rgb}{0.553,0.682,0.063}
\definecolor{RUBGRAY_RGB}{rgb}{0.906,0.906,0.906}
\definecolor{RUBGRAYDARK_RGB}{rgb}{0.588,0.588,0.588}

\definecolor{gray}{gray}{0.5}
\definecolor{gray70}{gray}{0.3}
\definecolor{gray60}{gray}{0.4}
\definecolor{gray40}{gray}{0.6}
\definecolor{gray30}{gray}{0.7}
\definecolor{gray20}{gray}{0.8}
\definecolor{gray10}{gray}{0.9}
\definecolor{gray05}{gray}{0.95}

\newcommand{\setsymbol}[1]{\ensuremath{\mathbb{#1}}}

\newcommand{\dsetcons}[2]%
{%
\ensuremath%
 {%
  \ifthenelse{\equal{#2}{0}}{[ #1 ]}{}%
  \ifthenelse{\equal{#2}{1}}{\bigl[ #1 \bigr]}{}%
  \ifthenelse{\equal{#2}{2}}{\Bigl[ #1 \Bigr]}{}%
  \ifthenelse{\equal{#2}{3}}{\biggl[ #1 \biggr]}{}%
  \ifthenelse{\equal{#2}{4}}{\Biggl[ #1 \Biggr]}{}%
 }%
}

\newcommand{\dsetconsnonneg}[2]%
{%
\ensuremath%
 {%
  \ifthenelse{\equal{#2}{0}}{[ #1 ]_{0}}{}%
  \ifthenelse{\equal{#2}{1}}{\bigl[ #1 \bigr]_{0}}{}%
  \ifthenelse{\equal{#2}{2}}{\Bigl[ #1 \Bigr]_{0}}{}%
  \ifthenelse{\equal{#2}{3}}{\biggl[ #1 \biggr]_{0}}{}%
  \ifthenelse{\equal{#2}{4}}{\Biggl[ #1 \Biggr]_{0}}{}%
 }%
}

\newcommand{\vect}[1]{\ensuremath{\mathbf{#1}}}

\newcommand{\transsymbol}{\ensuremath{\mathrm{T}}}
\newcommand{\trans}[1]{\ensuremath{#1^{\transsymbol}}}

\newcommand{\diagsymbol}{\ensuremath{\mathrm{diag}}}

\newcommand{\ddiag}[2]%
{%
 \ensuremath
 {%
  \ifthenelse{\equal{#2}{0}}{\diagsymbol\{ #1 \}}{}%
  \ifthenelse{\equal{#2}{1}}{\diagsymbol\bigl\{ #1 \bigr\}}{}%
  \ifthenelse{\equal{#2}{2}}{\diagsymbol\Bigl\{ #1 \Bigr\}}{}%
  \ifthenelse{\equal{#2}{3}}{\diagsymbol\biggl\{ #1\biggr \}}{}%
  \ifthenelse{\equal{#2}{4}}{\diagsymbol\Biggl\{ #1\Biggr \}}{}%
 }%
}

\newcommand{\dabs}[2]
{%
 \ensuremath%
 {
  \ifthenelse{\equal{#2}{0}}{\lvert#1\rvert}{}%
  \ifthenelse{\equal{#2}{1}}{\bigl\lvert#1\bigr\rvert}{}%
  \ifthenelse{\equal{#2}{2}}{\Bigl\lvert#1\Bigr\rvert}{}%
  \ifthenelse{\equal{#2}{3}}{\biggl\lvert#1\biggr\rvert}{}%
  \ifthenelse{\equal{#2}{4}}{\Biggl\lvert#1\Biggr\rvert}{}%
 }%
}

\newcommand{\dinprod}[3]{%
\ensuremath%
 {%
  \ifthenelse{\equal{#3}{0}}{\langle #1, #2 \rangle}{}%
  \ifthenelse{\equal{#3}{1}}{\bigl\langle #1, #2 \bigr\rangle}{}%
  \ifthenelse{\equal{#3}{2}}{\Bigl\langle #1, #2 \Bigr\rangle}{}%
  \ifthenelse{\equal{#3}{3}}{\biggl\langle #1, #2 \biggr\rangle}{}%
  \ifthenelse{\equal{#3}{4}}{\Biggl\langle #1, #2 \Biggr\rangle}{}%
 }%
}
\newcommand{\dinprodr}[3]{%
\ensuremath%
 {%
  \ifthenelse{\equal{#3}{0}}{\langle #1, #2 \rangle_{\text{r}}}{}%
  \ifthenelse{\equal{#3}{1}}{\bigl\langle #1, #2 \bigr\rangle_{\text{r}}}{}%
  \ifthenelse{\equal{#3}{2}}{\Bigl\langle #1, #2 \Bigr\rangle_{\text{r}}}{}%
  \ifthenelse{\equal{#3}{3}}{\biggl\langle #1, #2 \biggr\rangle_{\text{r}}}{}%
  \ifthenelse{\equal{#3}{4}}{\Biggl\langle #1, #2 \Biggr\rangle_{\text{r}}}{}%
 }%
}

\newcommand{\term}[1]{``#1''}

\acrodef{AAPM}{American Association of Physicists in Medicine}		
\acrodef{ACB}{asyn\-chro\-nous compressed beamformer}			
\acrodef{ACA}{adaptive cross approximation}				
\acrodef{ACR}{American College of Radiology}				
\acrodef{ADC}{analog-to-digital conversion}				
\acrodef{ADMM}{alternating direction method of multipliers}             
\acrodef{AIC}{analog-to-information conversion}				
\acrodef{BIBO}{bounded-input, bounded-output}				
\acrodef{BIM}{\name{Born} iterative method}				
\acrodef{BP}{basis pursuit}						

\acrodef{BPDN}{basis pursuit denoise}					
\acrodef{BVP}{boundary value problem}					
\acrodef{CAR}{Canadian Association of Radiologists}			
\acrodef{CDF}{cumulative distribution function}				
\acrodef{CGP}{conjugate gradient pursuit}				
\acrodef{CNR}{contrast-to-noise ratio}					
\acrodef{CoSaMP}{compressive sampling matching pursuit}			
\acrodef{CPU}{central processing unit}					
\acrodef{CPV}{\name{Cauchy} principal value}				
\acrodef{CPWC}{coherent plane-wave compounding}				

\acrodef{CS}{compressed sensing}					
\acrodef{CSI}{contrast source inversion}				
\acrodef{CT}{x-ray computed tomography}					
\acrodef{CTTE}[CT-TE]{continuous-time ternary encoding}			
\acrodef{CW}{cylindrical wave}						
\acrodef{DAC}{digital-to-analog conversion}				
\acrodef{DAIC}{diagnostically acceptable irreversible compression}	
\acrodef{DAS}{delay-and-sum}						
\acrodef{DBIM}{distorted \name{Born} iterative method}			
\acrodef{DCT}{discrete cosine transform}				

\acrodef{DICOM}{Digital Imaging and Communications in Medicine}		
\acrodef{DFT}{discrete \name{Fourier} transform}			
\acrodef{DOF}{depth of field}			
\acrodef{DSFT}{discrete space \name{Fourier} transform}			
\acrodef{DWT}{discrete wavelet transform}				
\acrodef{DNA}{deoxyribonucleic acid}					
\acrodef{DRG}{German \name{Röntgen} Society}				
\acrodef{DRIM}{distorted \name{Rytov} iterative method}			
\acrodef{DRS}{digital random sampler}					
\acrodef{ERM}{exploding reflector model}				

\acrodef{ESR}{European Society of Radiology}				
\acrodef{EU}{European Union}						
\acrodef{FBP}{filtered backpropagation}					
\acrodef{FDA}{US Food and Drug Administration}				
\acrodef{FDBF}{frequency domain beamforming}				
\acrodef{FDT}{\name{Fourier} diffraction theorem}			
\acrodef{FEHM}{full extent at half maximum}				
\acrodefplural{FEHM}{full extents at half maximum}			
\acrodef{FFT}{fast \name{Fourier} transform}				
\acrodef{FIR}{finite impulse response}					
\acrodef{FISTA}{fast iterative shrinkage-thresholding algorithm}        

\acrodef{FLOP}{floating point operation}				
\acrodef{FMM}{fast multipole method}					
\acrodef{FOCUS}{fast object-oriented C++ ultrasound simulator}		
\acrodef{FOV}{field of view}						
\acrodefplural{FOV}{fields of view}					
\acrodef{FPDE}{fractional partial differential equation}		
\acrodef{FPS}{frames per second}					
\acrodef{FRI}{finite rate of innovation}				
\acrodef{FWHM}{full width at half maximum}				
\acrodefplural{FWHM}{full widths at half maximum}			
\acrodef{FWT}{fast wavelet transform}					
\acrodef{GCNR}[gCNR]{generalized contrast-to-noise ratio}		

\acrodef{GLL}{grating lobe level}					
\acrodef{GPU}{graphics processing unit}					
\acrodef{GWN}{Gaussian white noise}					
\acrodef{HSV}{hue, saturation, value}					
\acrodef{HTP}{hard thresholding pursuit}				
\acrodef{IQR}{interquartile range}		        		

\acrodef{IHT}{iterative hard thresholding}				
\acrodef{IID}[i.i.d.]{independent and identically distributed}		
\acrodef{IQ}{in-phase and quadrature}					
\acrodef{IRLS}{iteratively reweighted least squares}			
\acrodef{ISP}{inverse scattering problem}				

\acrodef{ISR}{integrated sidelobe ratio}				
\acrodef{JPEG}{Joint Photographic Experts Group}			
\acrodef{JSM}{joint sparsity model}					
\acrodef{KK}{\name{Kramers}-\name{Kronig}}                              
\acrodef{LASSO}{least absolute shrinkage and selection operator}	
\acrodef{LDT}{\name{Laplace} diffraction theorem}			
\acrodef{LS}{\name{Lippmann}-\name{Schwinger}}				
\acrodef{LSB}{least significant bit}					
\acrodef{LTI}{linear time-shift invariant}				
\acrodef{MAP}{maximum \emph{a posteriori}}				

\acrodef{MCMC}{\name{Markov} chain Monte Carlo}				
\acrodef{MIRF}{material impulse response function}			
\acrodef{MLFMA}{multilevel fast multipole algorithm}			
\acrodef{MMV}{multiple measurement vectors}				
\acrodef{MOM}[MoM]{method of moments}					
\acrodef{MP}{matching pursuit}						
\acrodef{MP3}{MPEG Audio Layer III}					
\acrodef{MRI}{magnetic resonance imaging}				
\acrodef{MTF}{material transfer function}				
\acrodef{MV}{minimum variance}						

\acrodef{MVDR}{minimum-variance distortionless response}		
\acrodef{MWC}{modulated wideband converter}				
\acrodef{NESTA}{\name{Nesterov}'s algorithm}				
\acrodef{NSP}{null space property}					
\acrodef{NNZC}{number of nonzero components}				
\acrodef{OMP}{orthogonal matching pursuit}				
\acrodef{PAM}{pulse-amplitude modulation}				
\acrodef{PAT}{photoacoustic tomography}					
\acrodef{PC}{personal computer}						
\acrodef{PCIe}{peripheral component interconnect express}               
\acrodef{PDE}{partial differential equation}				

\acrodef{PDF}{probability density function}				
\acrodef{PET}{positron emission tomography}				
\acrodef{PICMUS}{Plane-Wave Imaging Challenge in Medical Ultrasound}	
\acrodef{PSF}{point spread function}					
\acrodef{PSL}{peak sidelobe level}					
\acrodef{PSNR}{peak signal-to-noise ratio}				
\acrodef{PW}{plane wave}						
\acrodef{QCW}{quasi-cylindrical wave}					
\acrodef{QPW}{quasi-plane wave}						
\acrodef{QSW}{quasi-spherical wave}					

\acrodef{QQ}{quantile-quantile}						
\acrodef{RA}{randomly-apodized}						
\acrodef{RAD}{randomly-apodized and randomly-delayed}			
\acrodef{RADAR}[RADAR]{radio detection and ranging}			
\acrodef{RAM}{random-access memory}					
\acrodef{RCR}{Royal College of Radiologists}				
\acrodef{RD}{randomly-delayed}						
\acrodef{RF}{radio frequency}						
\acrodef{RIC}{restricted isometry constant}				
\acrodef{RIP}{restricted isometry property}				

\acrodef{RMSE}{root mean-squared error}					
\acrodef{ROI}{region of interest}					
\acrodefplural{ROI}{regions of interest}				
\acrodef{ROMP}{regularized \ac{OMP}}					
\acrodef{RS}{\name{Rayleigh}-\name{Sommerfeld}}                         
\acrodef{SaS}[S$\alpha$S]{symmetric $\alpha$-stable}			
\acrodef{SA}{synthetic aperture}					
\acrodef{SBL}{sparse Bayesian learning}					
\acrodef{SDK}{software development kit}					
\acrodef{SDR}{software-defined radio}					
\acrodef{SIIM}{Society for Imaging Informatics in Medicine}		

\acrodef{SISO}{single-input, single-output}				
\acrodef{SINR}{signal-to-interference-plus-noise ratio}			
\acrodef{SMI}{sample matrix inversion}					
\acrodef{SNR}{signal-to-noise ratio}					
\acrodef{SOMP}{simultaneous orthogonal matching pursuit}		
\acrodef{SoS}{sum of sincs}						
\acrodef{SP}{subspace pursuit}						
\acrodef{spark}{sparse rank}						
\acrodef{SPECT}{single photon emission computed tomography}		
\acrodef{SPG}{spectral projected gradient}				

\acrodef{SPGL1}[SPG$\ell_{1}$]{spectral projected gradient for $\ell_{1}$-minimization}	
\acrodef{SPLS}{split-projection least squares}				
\acrodef{SR}{sparse recovery}						
\acrodef{SRC}{\name{Sommerfeld} radiation condition}			
\acrodef{SSI}{supersonic shear imaging}					
\acrodef{SSIM}{structural similarity}					
\acrodef{STFT}{short-time \name{Fourier} transform}			
\acrodef{StOMP}{stagewise \ac{OMP}}					
\acrodef{StWGP}{stagewise weak \ac{CGP}}				
\acrodef{SVD}{singular value decomposition}				

\acrodef{TDMA}{time-division multiple access}				
\acrodef{TMSBL}[T-MSBL]{temporally correlated \ac{MMV}-based \ac{SBL}}  
\acrodef{TOF}{time-of-flight}						
\acrodefplural{TOF}{times-of-flight}					
\acrodef{TGC}{time gain compensation}					
\acrodef{THI}{tissue harmonic imaging}					
\acrodef{TPSF}{transform point spread function}				
\acrodef{TPU}{tensor processing unit}                                   
\acrodef{TR}{\name{Tikhonov} regularization}				
\acrodef{TV}{total variation}						
\acrodef{UCA}{ultrasound contrast agents}				

\acrodef{UDT}{ultrasound diffraction tomography}			
\acrodef{UI}{ultrasound imaging}					
\acrodef{UoS}{union of subspaces}					
\acrodef{USA}{United States of America}					
\acrodef{UWB}{ultra-wideband}						
\acrodef{voxel}{volume element}						
\acrodef{Xampling}{CS-Sampling}						

\acresetall

\begin{document}

\pgfmathsetmacro{\xLB}{ -19.431 }
\pgfmathsetmacro{\zLB}{ 4.9 }
\pgfmathsetmacro{\zUB}{ 43.9 }

\pgfmathsetmacro{\FNumberTXConstant}{ 3.1 }
\pgfmathsetmacro{\FNumberTXProposedDOF}{ 13 }

\pgfmathsetmacro{\TXLengthZero}{ 11.4 }
\pgfmathsetmacro{\TXLengthOne}{ 24.4 }
\pgfmathsetmacro{\TXLengthTwo}{ 37.4 }

\pgfmathsetmacro{\FNumberRXConstant}{ 1.5 }

\pgfmathsetmacro{\IQRImprovementLB}{ 9.6 }
\pgfmathsetmacro{\IQRImprovementUB}{ 14.1 }
\pgfmathsetmacro{\GCNRImprovementLB}{ 6.9 }
\pgfmathsetmacro{\GCNRImprovementUB}{ 8.3 }

\pgfmathsetmacro{\FWHMMedianImprovementUB}{ 15.9 }
\pgfmathsetmacro{\FWHMSingleImprovementMag}{ 20.5 }
\pgfmathsetmacro{\FWHMSingleImprovementUB}{ 24 }

\begin{titlepage}
\thispagestyle{empty}%
\noindent
{\huge
Frequency-Dependent F-Numbers Suppress
Grating Lobes and\\[0.8ex] Improve
the Lateral Resolution in
Line-by-Line Scanning
}
\par
\vspace{36pt}
\noindent
{\large Martin F. Schiffner\par\vspace{12pt}
\noindent \href{http://www.mt.rub.de}{Chair of Medical Engineering}, Ruhr University Bochum, 44780 Bochum, Germany}
\vspace{36pt}
\par
\noindent
{\bf Copyright notice:}\par\vspace{12pt}
\noindent
\copyright~2024~IEEE.
Personal use of
this material is
permitted.
Permission from
IEEE must be obtained for
all other uses, in
any current or
future media, including reprinting/republishing
this material for
advertising or
promotional purposes, creating
new collective works, for
resale or
redistribution to
servers or
lists, or
reuse of
any copyrighted component of
this work in
other works.
\par
\vspace{12pt}
\noindent
{\bf Full citation:}\par\vspace{12pt}
\noindent
2024 IEEE Ultrason., Ferroelectr., and Freq. Control Joint Symp. (UFFC-JS), Taipei, Taiwan, Sep. 2024, pp. 1--4.
\par
\noindent
DOI: \href{https://doi.org/10.1109/UFFC-JS60046.2024.10793642}{10.1109/UFFC-JS60046.2024.10793642}
\par
\vspace{12pt}
\noindent
\href{https://ieeexplore.ieee.org/document/10793642}{Click here for IEEE Xplore}
\clearpage
\end{titlepage}

\title{Frequency-Dependent F-Numbers Suppress Grating Lobes and Improve the Lateral Resolution in Line-by-Line Scanning}

\author{%
  \IEEEauthorblockN{Martin F. Schiffner}%
  \IEEEauthorblockA{%
    \textit{Chair of Medical Engineering}\\
    \textit{Ruhr University Bochum}\\
    44780 Bochum, Germany\\
    0000-0002-4896-2757\,\orcidlink{0000-0002-4896-2757}
  }
}

\maketitle

\begin{abstract}
Line-by-line scanning with
linear arrays is
a standard image formation method in
clinical ultrasound.
This method examines progressively
a given \acl{ROI} by conducting
focused pulse-echo measurements with
dynamic transmit and
receive apertures.
Such apertures widen with
the focal length as
a function of
a given F-number and improve
the image quality by extending
the \ac{DOF} and suppressing
grating lobes.
Fixed F-numbers, however, limit
the lateral resolution.
Herein,
frequency dependence of
the F-number is incorporated into both
the transmit and
the receive focusing to widen
the apertures for
low frequencies and improve
the lateral resolution.
Frequency-dependent transmit and
receive F-numbers are
proposed.
These F-numbers, which can be
expressed in
closed form, maximize
the lateral resolution under constraints on
the \ac{DOF} and
the grating lobes.
A phantom experiment showed that
the proposed F-numbers eliminate
grating lobe artifacts and improve both
image uniformity and
contrast to
a similar extent as
fixed F-numbers.
These metrics, compared to
the usage of
the full apertures, improved by
up to \SI{\IQRImprovementUB}{\percent} and
\SI{\GCNRImprovementUB}{\percent},
respectively.
The proposed F-numbers, however, improved
the lateral resolution by
up to \SI{\FWHMSingleImprovementUB}{\percent} compared to
the fixed F-numbers.

\end{abstract}
\acresetall

\begin{IEEEkeywords}
  beamforming,
  dynamic aperture,
  F-number,
  frequency dependence,
  line-by-line scanning,
  progressive scanning
\end{IEEEkeywords}

\section{Introduction}
\label{sec:introduction}
Line-by-line scanning with
linear arrays is
a standard in
clinical ultrasound imaging
\cite{article:IlovitshITUFFC2019}.
This standard examines progressively
a given \ac{ROI} in
soft tissues by conducting
a sequence of
pulse-echo measurements.
Each measurement provides only
a single image line and begins with
the emission of
a focused wave.
This wave converges at
a transmit focus
$\vect{r}_{\text{f}}^{(n)} = \trans{ ( x_{\text{f}}^{(n)}, z_{\text{f}} ) }$.
The lateral focal position
$x_{\text{f}}^{(n)}$ increases with
the measurement index
$n$, whereas
the focal length
$z_{\text{f}}$ remains
constant during
all measurements.
A focused signal, whose envelope equals
the image line, then results from
dynamic receive focusing.
The receive focus tracks
the emitted wave along
the line
$x = x_{\text{f}}^{(n)}$. 
The image quality, however, degrades away from
the transmit focal length
$z_{\text{f}}$.
This degradation motivates
methods to approximate
dynamic transmit focusing
\cite{article:IlovitshITUFFC2019}.
One of these methods is
compounding of
multiple images with
different transmit focal lengths.
Such compounding improves
the image quality at
the expense of
the temporal resolution.

The focusing uses
a technique known as
\term{dynamic aperture} to improve
image uniformity and suppress
grating lobes at
the expense of
the lateral resolution
\cite{%
  article:SchiffnerUlt2024,
  article:SchiffnerITUFFC2023,
  proc:SchiffnerIUS2021,
  article:WilcoxITUFFC2018,
  article:DelannoyJAP1979%
}.
Grating lobes result from
violation of
the sampling theorem by
the linear arrays and, usually, cause
image artifacts.
Linear arrays, in fact, use
a relatively large element pitch
$p$, which approximates
the center wavelength
$\lambda_{\text{c}}$
(i.e., $p \approx \lambda_{\text{c}}$), to widen
the \ac{FOV} near
the skin surface.
The dynamic aperture ensures that
the focusing, for
any given focus, uses only
a specific set of
array elements.
This set is supposed to be
centered on
the lateral focal coordinate
$x_{\text{f}}^{(n)}$.
The desired aperture width
$A( z_{\text{f}} )$ increases linearly with
the focal length
$z_{\text{f}}$ as
a function of
a user-defined F-number
\cite[(1)]{article:SchiffnerITUFFC2023},
\cite[173]{book:Cobbold2006},
\cite[(12)]{article:DelannoyJAP1979}
\begin{equation}
  F
  =
  \frac{
    z_{\text{f}}
  }{
    A( z_{\text{f}} )
  },
 \label{eqn:theory_f_number}
\end{equation}
which usually ranges from
\numrange{1}{3}.
The F-number, as will be shown in
\cref{subsec:theory_beam_properties}, determines
essential properties of
the focused wave.
Recent results in
ultrafast plane-wave imaging
\cite{%
  article:SchiffnerITUFFC2023,
  proc:SchiffnerIUS2021%
}, moreover, suggest that
the F-number
\eqref{eqn:theory_f_number} should increase monotonically with
the frequency
$f$.
Such frequency dependence describes
a dynamic aperture that not only varies with
the focus but also narrows with
the frequency
$f$ and, to
the best knowledge of
the author, has never been incorporated into
line-by-line scanning.

Herein,
frequency dependence of
the F-number
\eqref{eqn:theory_f_number} is incorporated into both
the transmit and
the receive focusing of
line-by-line scanning.
The effects of
the F-number
\eqref{eqn:theory_f_number} and
the frequency
$f$ on
the focusing will be investigated first.
Subsequently,
closed-form expressions for
frequency-dependent transmit and
receive F-numbers will be
proposed.
Both F-numbers attempt to maximize
the lateral resolution under
constraints.
The proposed transmit F-number prevents
the image degradation away from
the transmit focal length
$z_{\text{f}}$.
The proposed receive F-number, in contrast, limits
the grating lobe amplitudes as in
\cite{article:SchiffnerITUFFC2023}.
The effects of
both F-numbers will be investigated separately in
experiments on
a tissue phantom.
The resulting images show that
the proposed F-numbers improve
image uniformity and reduce
artifacts to
a similar extent as
fixed F-numbers but improve significantly
the lateral resolution.

\section{Theory}
This paper exclusively treats
the compounding of
multiple images with
different transmit focal lengths.
The proposed F-numbers will be
presented after investigating
the effects of
the F-number
\eqref{eqn:theory_f_number} and
the frequency
$f$ on
the focusing.
These effects were
determined by simulating
a commercial linear array, which was used in
the experiments
(see \cref{sec:methods}), in
Field II
\cite{%
  article:JensenMBEC96,
  article:JensenITUFFC1992%
}.

\subsection{Effects of the F-Number and the Frequency on the Focusing}
\label{subsec:theory_beam_properties}
%
\begin{figure}[t!]
 \centering%
  \input{theory/figures/latex/beam_properties.tex}
 \caption{}
 \label{fig:V}
\end{figure}
C
{
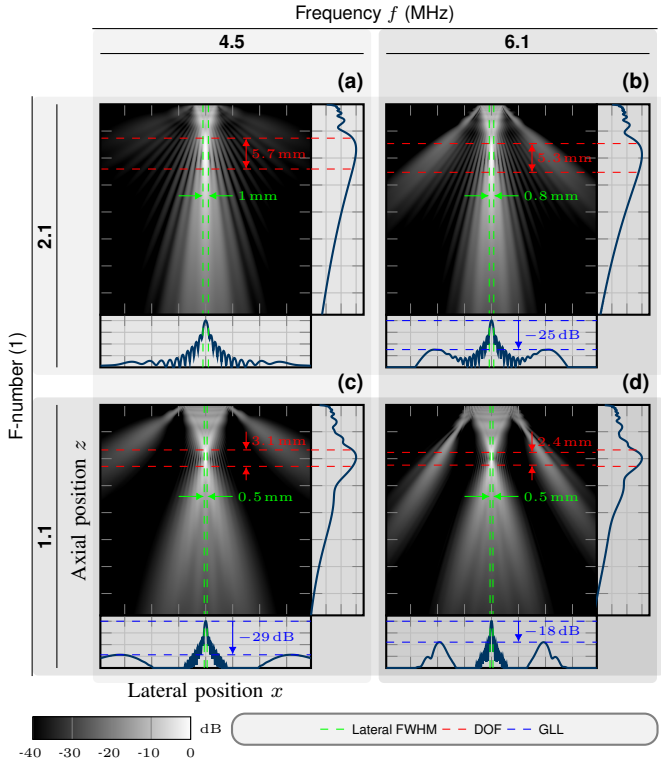
 Effects of
 the F-number
 \eqref{eqn:theory_f_number} and
 the frequency
 $f$ on
 the focused beam.
 A decrease in
 the F-number
 \eqref{eqn:theory_f_number} or 
 an increase in
 the frequency
 $f$ has
 the following effects:
 1)
 the lateral \acs{FWHM} and
 the \acs{DOF} decrease; and
 2)
 the \acs{GLL} increases.
 Frequency dependence of
 the F-number
 \eqref{eqn:theory_f_number}, hence, can keep constant
 a selected metric, such as
 the \acs{DOF} or
 the \acs{GLL}.
 The images show
 the focused beam at
 the center frequency
 (left column) and
 near the upper frequency bound
 (right column) with
 the lateral and
 axial profiles for
 a large F-number (top row) and
 a small F-number (bottom row).
 The axes in
 all images are
 equal.
}%
{theory_beams}

The F-number
\eqref{eqn:theory_f_number} and
the frequency
$f$, as shown in
\cref{fig:theory_beams}, determine
essential properties of
the focused beam.
These properties include:
1)
the lateral \ac{FWHM};
2)
the \ac{DOF}; and
3)
the \ac{GLL}.
As
the F-number
\eqref{eqn:theory_f_number} decreases or
the frequency
$f$ increases,
the lateral \ac{FWHM} and
the \ac{DOF} decrease.
Both properties indicate
ranges of
positions over which
the beam is
focused satisfactorily.
The lateral \ac{FWHM} equals
the distance between
the lateral positions in
the focal plane where
the beam intensity reduces by
\SI{6}{\deci\bel}
\cite[p. 173]{book:Cobbold2006}.
Smaller
lateral \acp{FWHM} indicate
better lateral resolution.
The \ac{DOF}, similarly, equals
the distance between
the axial positions where
the beam intensity reduces by
\SI{2.2}{\deci\bel}
\cite[p. 173]{book:Cobbold2006},
\cite[p. 491]{book:Born1999}.
This property, owing to
the few transmit focal lengths
$z_{\text{f}}$ in
a compound image
(see \cref{sec:introduction}), is only relevant to
the transmit focusing.
Larger
\acp{DOF} reduce
image degradation away from
the transmit focal length
$z_{\text{f}}$.
The \ac{GLL}, in contrast, increases as
the F-number
\eqref{eqn:theory_f_number} decreases or
the frequency
$f$ increases.
The \ac{GLL} denotes
the ratio of
the maximum amplitudes attained by
the grating lobes and
the main lobe
\cite{%
  article:SchiffnerITUFFC2023,
  article:DelannoyJAP1979%
} and indicates
the dynamic range available for
unambiguous imaging.
Smaller
\acp{GLL} reduce
grating lobe artifacts.

\subsection{Proposed Transmit F-Number}
\label{subsec:theory_f_number_transmit}
A transmit F-number will now be proposed.
This F-number attempts to maximize
the lateral resolution while preventing
image degradation away from
the transmit focal length
$z_{\text{f}}$.
This objective is achieved by maintaining
a constant \ac{DOF} for
all relevant frequencies.
The desired \ac{DOF} derives from
the axial length of
the \ac{FOV} and
the number of
transmit focal lengths
$N_{\text{foc}, z}$ in
a compound image
(i.e.,
  $\text{\acs{DOF}} = ( z_{\text{ub}} - z_{\text{lb}} ) / N_{\text{foc}, z}$, where
  $z_{\text{lb}}$ and
  $z_{\text{ub}}$ denote
  the lower and
  upper bounds on
  the axial position%
).
The \ac{DOF}, moreover, is
proportional to
the product of
the wavelength
$\lambda$ and
the square of
the F-number
\eqref{eqn:theory_f_number}
\cite[p. 491]{book:Born1999}:
\begin{equation*}
  \text{\acs{DOF}}
  \approx
  6.1 \lambda F^{2}.
\end{equation*}
The proposed transmit F-number, hence, reads
\begin{equation}
  F( \lambda )
  =
  \sqrt{
    \frac{
      \text{\acs{DOF}}
    }{
      6.1 \lambda
    }
  }
 \label{eqn:theory_f_number_tx}
\end{equation}
and, owing to
the relation
$\lambda = c / f$ with
the average speed of sound
$c$, increases with
the square root of
the frequency
$f$.

\subsection{Proposed Receive F-Number}
The receive F-number equals
the F-number proposed in
\cite{article:SchiffnerITUFFC2023}.
This F-number attempts to maximize
the lateral resolution under
two constraints on
the grating lobes.
These constraints, whose details are
outside the scope of
this article, limit
the \ac{GLL} by:
1)
avoiding
lobe aliasing; and
2)
imposing
a minimum angular distance
$\chi_{0}$ on
the first-order grating lobes.
The proposed receive F-number is
a function of
the normalized element pitch
$\bar{p} = p / \lambda$ and reads
\cite[(21)]{article:SchiffnerITUFFC2023}
\begin{subequations}
\label{eqn:theory_f_number_rx}
\begin{equation}
  F( \bar{p} )
  =
  \max\left\{
    F_{\text{lb}}^{(\text{A})}( \bar{p} ),
    F_{\text{lb}}^{(\text{G})}( \bar{p} )
  \right\},
 \label{eqn:theory_f_number_rx_max}
\end{equation}
where
\begin{equation}
  F_{\text{lb}}^{(\text{A})}( \bar{p} )
  =
  \begin{cases}
    0
  & \text{for } \bar{p} < 0.5,\\
    \frac{
      \sqrt{ { \bar{p} }^{2} - 0.25 }
      +
      { \bar{p} }^{2} \delta
    }{
      1 - { \bar{p} }^{2} \delta^{2}
    }
  & \text{for } 0.5 \leq \bar{p} < 1 / \delta,\\
  \end{cases}
 \label{eqn:theory_f_number_lobe_anti_aliasing}
\end{equation}
separates
the first-order grating lobes from
the main lobe by
an angle
$\delta$ and
\begin{equation}
  F_{\text{lb}}^{(\text{G})}( \bar{p} )
  =
  \begin{cases}
    0
  & \text{for } \bar{p} \leq \bar{p}_{\text{lb}},\\
    \frac{
      1
    }{
      2
    }
    \sqrt{
      \frac{
        1
      }{
        \left[
          1 / \bar{p}
          -
          \sin( \chi_{0} )
        \right]^{2}
      }
      -
      1
    }
  & \text{for } \bar{p} \in \setsymbol{P},\\
    F_{\text{ub}}
  & \text{for } \bar{p} \geq \bar{p}_{\text{ub}},\\
  \end{cases}
\end{equation}
with
the set
$\setsymbol{P} = ( \bar{p}_{\text{lb}}; \bar{p}_{\text{ub}} )$ and
the bounds
\begin{align}
  \bar{p}_{\text{lb}}
  &=
  \frac{
    1
  }{
    \sin( \chi_{0} ) + 1
  },
 \label{eqn:theory_f_number_norm_element_pitch_lb}\\
  \bar{p}_{\text{ub}}
  &=
  \frac{
    1
  }{
    \sin( \chi_{0} )
    +
    \frac{
      1
    }{
      \sqrt{ 1 + ( 2 F_{\text{ub}} )^{2} }
    }
  }.
 \label{eqn:theory_f_number_norm_element_pitch_ub}
\end{align}
\end{subequations}
ensures
the minimum angular distance
$\chi_{0}$.
The maximum permissible F-number
$F_{\text{ub}}$ avoids
very narrow apertures to sustain
the focusing.
\TODO{typical choice}
The reader is referred to
\cite{article:SchiffnerITUFFC2023} for
any details.

\section{Methods}
\label{sec:methods}
The advantages of
dynamic apertures over
the full apertures and
the proposed F-numbers
\eqref{eqn:theory_f_number_tx} and
\eqref{eqn:theory_f_number_rx} over
the usual fixed F-numbers were confirmed in
an experiment with
a commercial multi-tissue phantom%
\footnote{%
  Computerized Imaging Reference Systems (CIRS), Inc., Norfolk, VA, USA%
  \label{ftnote:manufacturer_cirs}
}
(%
  model: 040;
  average speed of sound: $c = \SI{1538.75}{\meter\per\second}$%
).
A SonixTouch Research system%
\footnote{%
  Analogic Corporation, Sonix Design Center, Richmond, BC, Canada
  \label{ftnote:manufacturer_analogic}
} with
a linear array
(%
  model: L14-5/38;
  number of elements: \num{128},
  element width: \SI{279.8}{\micro\meter},
  element height: \SI{4}{\milli\meter},
  pitch: $p = \SI{304.8}{\micro\meter}$,
  elevational focus: \SI{16}{\milli\meter}%
) acquired and stored
the \ac{RF} signals of
a complete synthetic aperture scan
\cite{article:JensenUlt2006} for
offline processing.
The excitation voltage was
a single cycle at
\SI{4}{\mega\hertz}.

\subsection{Image Formation}
Line-by-line scans were synthesized from
the acquired \ac{RF} signals in
the Fourier domain
\cite{article:BottenusITUFFC2018} for
three transmit focal lengths
$z_{\text{f}} \in \{ \SI{\TXLengthZero}{\milli\meter}, \SI{\TXLengthOne}{\milli\meter}, \SI{\TXLengthTwo}{\milli\meter} \}$
(i.e., $N_{\text{foc}, z} = 3$). 
The lower and
upper frequency bounds were
$f_{\text{lb}} = \SI{2.25}{\mega\hertz}$ and
$f_{\text{ub}} = \SI{6.75}{\mega\hertz}$,
respectively.
The apodization weights derived from
Tukey windows with
a cosine fraction of
\SI{20}{\percent}.
The number of
pulse-echo measurements, for each
transmit focal length
$z_{\text{f}}$, amounted to
$N = 256$, and
the transmit foci
$\vect{r}_{\text{f}}^{(n)} = \trans{ ( x_{\text{f}}^{(n)}, z_{\text{f}} ) }$ were equidistant
(i.e.,
  $x_{\text{f}}^{(n)} = \SI{\xLB}{\milli\meter} + n \SI{152.4}{\micro\meter}$ for
  all $0 \leq n < N$%
).
\TODO{512 focal lengths}
The receive focus, in
each measurement, had
the same lateral coordinate as
the transmit focus and tracked
the emitted wave along
the line
$x = x_{\text{f}}^{(n)}$.
The lower and
upper bounds on
the axial position were
$z_{\text{lb}} = \SI{\zLB}{\milli\meter}$ and
$z_{\text{ub}} = \SI{\zUB}{\milli\meter}$,
respectively.

\subsection{Investigated F-Numbers}
The transmit F-numbers were
the proposed F-number
\eqref{eqn:theory_f_number_tx} and
the fixed value of
this F-number at
the upper frequency bound
$f_{\text{ub}}$.
The required \ac{DOF} amounted to
$\text{\acs{DOF}} \approx \SI{\FNumberTXProposedDOF}{\milli\meter}$.
The fixed value at
the upper frequency bound was
$F \approx \FNumberTXConstant$.
The receive F-numbers were
the fixed F-number
$F = \FNumberRXConstant$ and
the proposed F-number
\eqref{eqn:theory_f_number_rx} with
$\chi_{0} = \SI{60}{\degree}$,
$F_{\text{ub}} = \FNumberRXConstant$, and
$\delta = \SI{10}{\degree}$.
The usage of
the full aperture in both
the transmit and
the receive focusing was enforced by setting
the F-numbers close to
zero.

\subsection{Image Post-Processing}
\label{subsec:methods_post_processing}
Compound images were formed by combining linearly
the three images for
all transmit focal lengths.
The coefficients in
these combinations varied with
the axial position and emphasized
the image whose
transmit focal length
$z_{\text{f}}$ was
closest to
the axial position
$z$.
The lateral \acp{FWHM} of
all wires and
the \ac{GCNR}
\cite{article:Rodriguez-MolaresITUFFC2020} of
the large anechoic region were
computed.
The author maintains
a public \name{Matlab}%
\footnote{%
  The MathWorks, Inc., Natick, MA, USA
  \label{ftnote:manufacturer_mathworks}
} source code
\cite{software:FNumber} to support
the reproduction of
the presented results and facilitate
further research.

\section{Results}
%
\begin{figure}[t!]
 \centering%
  \input{results/figures/latex/results_experiments_cirs_040_full_aperture.tex}
 \caption{}
 \label{fig:V}
\end{figure}
C
{
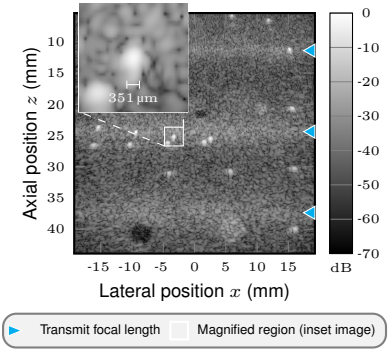
 Image of
 the multi-tissue phantom for
 the full apertures.
 These apertures achieved
 the best lateral resolution at
 the expense of
 grating lobe artifacts and
 image nonuniformity.
 The image shows
 the absolute voxel values for
 three transmit focal lengths
 (cyan triangles).
}%
{exp_val_cirs_040_images_full_aperture}

\begin{table}[tb!]
 \centering
 \caption{%
  Lateral \acsp{FWHM} of
  the wires and
  \acsp{GCNR} of
  the large anechoic region achieved by
  the full apertures and
  the dynamic apertures with
  all four combinations of
  investigated F-numbers.
 }
 \label{tab:metrics}
 \small
 \begin{tabular}{%
  @{}%
  l
  l
  S[table-format=3.0,table-number-alignment = right,table-auto-round,separate-uncertainty=true]
  S[table-format=3.0,table-number-alignment = right,table-auto-round,separate-uncertainty=true]
  S[table-format=2.1,table-number-alignment = right,table-auto-round]
  @{}%
 }
 \toprule
  \multicolumn{1}{@{}H}{\multirow{3}{*}{Transmit}} &
  \multicolumn{1}{H}{\multirow{3}{*}{Receive}} &
  \multicolumn{2}{H}{Lateral \acsp{FWHM}} &
  \multicolumn{1}{H@{}}{\multirow{2}{*}{\acs{GCNR}}}\\
  \cmidrule(lr){3-4}
  &
  &
  \multicolumn{1}{H}{Median} &
  \multicolumn{1}{H}{\acs{IQR}}\\
  &
  &
  \multicolumn{1}{H}{(\si{\micro\meter})} &
  \multicolumn{1}{H}{(\si{\micro\meter})} &
  \multicolumn{1}{H@{}}{(\si{\percent})}\\
  \cmidrule(r){1-1}\cmidrule(lr){2-2}\cmidrule(lr){3-3}\cmidrule(lr){4-4}\cmidrule(l){5-5}
 \addlinespace
  Full aperture & Full aperture & 493.39 & 177.16 & 87.4300\\ 
  Fixed         & Fixed         & 742.95 & 158.12 & 95.5628\\ 
  Proposed      & Fixed         & 727.71 & 158.12 & 95.6577\\ 
  Fixed         & Proposed      & 632.46 & 160.02 & 94.3142\\ 
  Proposed      & Proposed      & 624.84 & 152.40 & 95.3295\\ 
 \addlinespace
 \bottomrule
 \end{tabular}
\end{table}

The usage of
the full apertures, as shown in
\cref{fig:exp_val_cirs_040_images_full_aperture}, resulted in
grating lobe artifacts close to
the linear array and
nonuniform image quality.
The grating lobe artifacts resembled
moiré patterns
(i.e.,
  patterns of
  alternating dark and
  bright areas%
) that differred from
the usual speckle pattern
\cite{article:SchiffnerITUFFC2023}.
The wires close to
the transmit foci had
the smallest lateral \acsp{FWHM}.
These widths increased significantly away from
the transmit foci because
the \acp{DOF} were too small.
These small \acp{DOF}, as stated in
\cref{tab:metrics}, also resulted in
suboptimal contrast of
the large anechoic region.

%
\begin{figure}[t!]
 \centering%
  \input{results/figures/latex/results_experiments_cirs_040.tex}
 \caption{}
 \label{fig:V}
\end{figure}
C
{
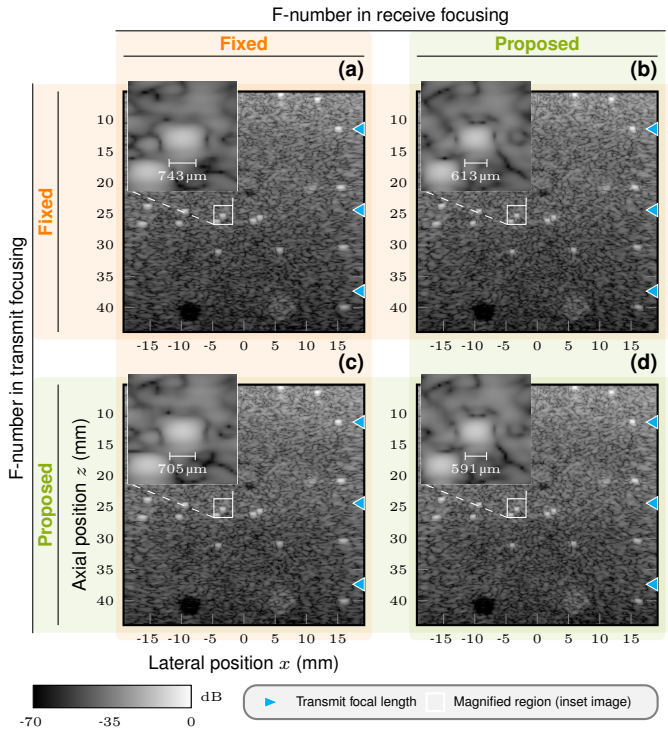
 Images of
 the multi-tissue phantom for
 dynamic apertures.
 All F-numbers, in comparison to
 the full apertures
 (see \cref{fig:exp_val_cirs_040_images_full_aperture}), eliminated
 grating lobe artifacts and improved both
 image uniformity and
 contrast.
 The proposed F-numbers, however, improved
 the lateral resolution by
 up to \SI{\FWHMSingleImprovementUB}{\percent} with respect to
 the fixed F-numbers.
 The images show
 the absolute voxel values for
 three transmit focal lengths
 (cyan triangles) and
 all combinations of
 transmit
 (rows) and
 receive F-numbers
 (columns).
 The transmit F-numbers are
 the fixed F-number
 $F \approx \FNumberTXConstant$
 (top row) and
 the proposed F-number
 \eqref{eqn:theory_f_number_tx} with
 $\text{DOF} \approx \SI{\FNumberTXProposedDOF}{\milli\meter}$
 (bottom row).
 The receive F-numbers are
 the fixed F-number
 $F = \FNumberRXConstant$
 (left column) and
 the proposed F-number
 \eqref{eqn:theory_f_number_rx} with
 $\chi_{0} = \SI{60}{\degree}$,
 $F_{\text{ub}} = \FNumberRXConstant$, and
 $\delta = \SI{10}{\degree}$
 (right column).
 The axes in
 all images are
 equal.
}%
{exp_val_cirs_040_images}

The usage of
dynamic apertures, as shown in
\cref{fig:exp_val_cirs_040_images}, eliminated
grating lobe artifacts and improved both
image uniformity and
contrast.
The \ac{IQR} of
the lateral \acp{FWHM} and
the \ac{GCNR} of
the large anechoic region, according to
\cref{tab:metrics}, improved by
up to \SI{\IQRImprovementUB}{\percent} and
\SI{\GCNRImprovementUB}{\percent},
respectively.
The exact lateral \acsp{FWHM} of
the wires, however, depended strongly on
the F-numbers.
The fixed F-numbers, as shown in
\cref{fig:exp_val_cirs_040_images}(a), increased
the lateral \acsp{FWHM} of
all wires in comparison to
the full apertures
(see \cref{fig:exp_val_cirs_040_images_full_aperture}).
This increase, as shown in
\cref{fig:exp_val_cirs_040_images}(c), was mitigated by
the proposed transmit F-number
\eqref{eqn:theory_f_number_tx}.
The proposed receive F-number
\eqref{eqn:theory_f_number_rx}, as shown in
\cref{fig:exp_val_cirs_040_images}(b), improved
this mitigation because
the dynamic receive focusing did not require
a large \ac{DOF}.
The combination of
the proposed F-numbers
\eqref{eqn:theory_f_number_tx} and
\eqref{eqn:theory_f_number_rx}, as shown in
\cref{fig:exp_val_cirs_040_images}(d), achieved
the best lateral \acp{FWHM} of
all four investigated combinations.
The lateral \ac{FWHM} of
the wire in
the magnified region, for example, reduced by
\SI{\FWHMSingleImprovementMag}{\percent} in comparison to
the fixed F-numbers.
The highest reduction of
\SI{24}{\percent} was achieved for
the wire in
the lower right corner.
The median lateral \ac{FWHM} of
all wires, according to
\cref{tab:metrics}, reduced by
\SI{\FWHMMedianImprovementUB}{\percent}.

\section{Conclusion}
Dynamic apertures are
essential to
line-by-line scanning with
linear arrays.
Such apertures improve
image uniformity and suppress
grating lobes at
the expense of
the lateral resolution.
The user, however, must specify
suitable F-numbers.
All investigated F-numbers eliminated
grating lobe artifacts and improved both
image uniformity and
contrast by
up to \SI{\IQRImprovementUB}{\percent} and
\SI{\GCNRImprovementUB}{\percent},
respectively.
The proposed frequency-dependent F-numbers, in comparison to
the fixed F-numbers, improved significantly
the lateral resolution by
up to \SI{\FWHMSingleImprovementUB}{\percent}.
The median improvement amounted
up to \SI{\FWHMMedianImprovementUB}{\percent}.
These findings suggest that exploiting
frequency dependence of
the F-numbers is
beneficial and deserves
further research.
Details of
the theory, such as
(i)
the implementation of
the frequency-dependent transmit F-number,
(ii)
the determination of
optimal parameters, and
(iii)
the effects on
the \acl{SNR}, were left to
another publication.
Future research will optimize
the F-numbers and adapt
the theory to
other imaging modes.


\bibliographystyle{IEEEtran}


\end{document}